\documentclass[preprint,12pt]{elsarticle}

\usepackage{graphicx}
\usepackage{amsmath}
\usepackage[usenames]{color}
\usepackage[breaklinks=true,colorlinks=true,linkcolor=blue,urlcolor=blue,citecolor=blue]{hyperref}
\usepackage{verbatim}
\usepackage{multirow}
\usepackage{braket}

\newcounter{bla}

\journal{Computer Physics Communications}

\begin{document}
	\graphicspath{{/Users/subashri/Dropbox/For-Subashri-Oleg-Rajesh/Aggregation_Rare_Events/01-Algorithm/Tex/Final/}}
	\begin{frontmatter}

		\title{A Monte Carlo algorithm to measure probabilities of rare events in cluster-cluster aggregation}

		\author[1]{Rahul Dandekar\corref{author1}}		
		\ead{rahul.dandekar@cea.fr}
		\author[2,3]{R. Rajesh\corref{author2}}
		\ead{rrajesh@imsc.res.in}
		\author[2,3]{V. Subashri\corref{author3}}
		\ead{subashriv@imsc.res.in}
		\author[4]{Oleg Zaboronski\corref{author4}}
		\ead{o.v.zaboronski@warwick.ac.uk}

		\address[1]{Institut de Physique Theorique, CEA, CNRS, Universite Paris–Saclay, F–91191 Gif-sur-Yvette cedex, France}
		\address[2]{The Institute of Mathematical Sciences, C.I.T. Campus, Taramani, Chennai 600113, India }
		\address[3]{Homi Bhabha National Institute, Training School Complex, Anushakti Nagar, Mumbai 400094, India}
		\address[4]{Mathematics Institute, University of Warwick, Coventry, CV4 7AL, United Kingdom}

		\begin{abstract}
		We develop a biased Monte Carlo algorithm to measure probabilities of rare events in cluster-cluster aggregation for arbitrary collision kernels. Given a trajectory with a fixed number of collisions, the algorithm modifies both the waiting times between collisions, as well as the sequence of collisions, using local moves. We show that the algorithm is ergodic by giving a protocol that transforms an arbitrary trajectory to a standard trajectory using valid Monte Carlo moves. The algorithm can sample rare events with probabilities of the order of $10^{-40}$ and lower. The algorithm's effectiveness in sampling low-probability events is established by showing that the numerical results for the large deviation function of constant-kernel aggregation reproduce the exact results. It is shown that the algorithm can obtain the large deviation functions for other kernels, including gelling ones, as well as the instanton trajectories for atypical times. The dependence of the autocorrelation times, both temporal and configurational,  on the different parameters of the algorithm is also characterized.
		\end{abstract}
		
			\end{frontmatter}
	\section{Introduction \label{intro}}

The study of the cluster-cluster aggregation  (CCA), in which particles, or clusters coalesce on contact to form larger clusters has a long history dating back to Smoluchowski in 1917~\cite{smoluchowski1917mathematical}. There are many physical phenomena in which the dominant dynamic process is coalescence or aggregation. Examples include blood coagulation~\cite{guria2009mathematical}, cloud formation~\cite{falkovich2002acceleration,pruppacher2012microphysics}, aerosol dynamics~\cite{williams1988unified},  dynamics of Saturn's rings~\cite{brilliantov2015size,connaughton2018stationary}, aggregation of particulate matter~\cite{burd2009particle} and rod-like phytoplanktons in oceans~\cite{slomka2020bursts}, protein aggregation~\cite{benjwal2006monitoring,wang2015following}, coagulation of soot particles~\cite{zhang2020three,liu2019numerical,sorensen2018light},  colloids~\cite{colomer2005experimental}, charged polymers~\cite{tom2016aggregation,tom2017aggregation}, etc. In addition to direct application, CCA is also of interest as a nonequilibrium process obeying self-similar dynamics with exponents that are universal and dependent only on generic details of the transport. These universal features has seen CCA being applied in seemingly unrelated areas  like Burgers turbulence~\cite{kida1979asymptotic,frachebourg2000ballistic,tribe2000large,dey2011lattice}, Kolmogorov self-similar scaling~\cite{takayasu1989steady,connaughton2005breakdown,connaughton2006cluster}, granular systems~\cite{ben1999shocklike,nie2002dynamics,pathak2014energy}, hydrodynamics of  run and tumble particles~\cite{dandekar2020hard}, evolution of planetesimals~\cite{wetherill1988formation}, geophysical flows~\cite{meakin1991fractal}, etc.

The most common approach to study CCA is to model it using the mean field Smoluchowski equation, an integro-differential equation for the rate of change of the number of clusters of a given size or mass. The transport process that brings clusters together is incorporated into the collision kernel that describes the rate of collision between  different masses. The most common transport processes are diffusive and ballistic transport. The Smoluchowski equation is exactly solvable for specific kernels, and a summary of results may be found in the following reviews~\cite{leyvraz2003scaling,aldous1999deterministic,krapivsky2010kinetic,handbook}.  
In lower dimensions, spatial fluctuations become important and may be studied using exact solution~\citep{spouge1988exact}, simulations~\cite{kang1984fluctuation}, or the renormalization group method~\citep{krishnamurthy2003persistence,krishnamurthy2002kang}. Irrespective of the approach, the primary focus has been on determining quantities that depend on typical events like the mean mass distribution, or its moments, like the mean number of particles, lower order moments of the mass distribution like number fluctuations, etc. To the best of our knowledge, there are no known results for the probabilities of atypical, rare events of CCA, the focus of this paper, where we propose an algorithm that can numerically determine these low probabilities in a model (Marcus-Lushnikov model)~\cite{connaughton2018stationary,nie2002dynamics,eibeck2001stochastic} that takes into account stochastic fluctuations but ignores spatial variations. 

Rare events are those which occur at the tails of a probability distribution, and have a low likelihood of occurrence. In a stochastic process, although rare events occur infrequently, they often have a large impact. Examples include cyclones, tsunamis, earthquakes~\cite{ben2020localization}, heat waves~\cite{ragone2021rare,ragone2018computation}, financial black swan events~\cite{morales2020covid19}, neurological disorders~\cite{iannucci2020thrombin} and pandemics like COVID-19. For predicting their occurrence in order to plan for them, it is important to have an estimate of the probability of occurrence as well as the atypical trajectories that lead to rare events.  Also, knowing the probabilities gives complete information about the large fluctuations of a system around its most probable states. The behaviour of the tails of the probability distributions describing these large fluctuations are captured by the large deviation function~\cite{touchette2009large}. The large deviation function is the central focus of study of large deviation theory, which has several physical applications. The large deviation function can be interpreted as a nonequilibrium generalization of entropy, using which, it can be proved that the scaled cumulant generating function associated with the distribution can be interpreted as a nonequilibrium generalization of free energy.

Numerically, many sophisticated techniques have been developed for studying rare events, such as importance sampling~\cite{hartmann2020probing,hartmann2011large,hartmann2018distribution} and splitting algorithms. In importance sampling, the original probability distribution is biased so that the rare event occurs more frequently.  The distribution is then unbiased to obtain the true probability of the event.  Different kinds of importance sampling methods have been developed to sample rare events, such as instanton based importance sampling~\cite{ebener2019instanton} and adaptive importance sampling~\cite{hartmann2019variational}. In splitting algorithms, events close to the rare event of interest are realized many times while other events are allowed with a certain probability, in the course of the simulation. Different types of splitting algorithms include static and dynamic splitting, and adaptive splitting algorithms~\cite{cerou2019adaptive}. A review of the different numerical methods available for calculating probability of rare events may be found in Refs.~\cite{bouchet2019rare,grafke2019numerical}.

The study of rare events in aggregation is  a challenging problem, because the number of possible configurations after each collision increases rapidly, which means that sampling these configurations would be a computationally expensive task. Is it possible to develop an algorithm which would be able to sample rare configurations robustly, and at the same time, be computationally efficient? Can we identify a large deviation rate function for the probability distribution obtained from such an algorithm?

In this paper, we develop a Monte Carlo algorithm to measure probabilities of rare events in CCA for arbitrary collision kernels. The algorithm is based on importance sampling. The key contribution in this  paper is to identify local modifications to a trajectory consistent with the collision rules, as well as the probabilities arising from collision rates and waiting times.  We show that the algorithm is ergodic by giving a protocol that transforms any given trajectory to a standard trajectory using reversible moves. The algorithm's effectiveness in sampling low-probability events is established by numerically reproducing the exact large deviation function for the constant-kernel aggregation. Further, it is shown that the algorithm can obtain the rate functions for gelling kernels, as well as the instanton trajectories for both typical and atypical times. The dependence of the autocorrelation times, both temporal and configurational, on the different parameters of the algorithm is also characterized.
 
The remainder of the paper is organized as follows. In Sec.~\ref{2}, the CCA model is defined. In Sec.~\ref{results}, the algorithm is described in detail. In Sec.~\ref{ergodicity}, we show that the algorithm is ergodic. In Sec.~\ref{ldev}, the probability distribution obtained from the algorithm is benchmarked with the exact answer, for constant kernel aggregation, where the rate of collision is independent of the colliding masses. A large deviation principle for arbitrary kernels is also identified numerically under a certain scaling limit. Section~\ref{typ} shows the typical as well as the rare trajectories for three different kernels. In Section~\ref{autocorr}, the behavior of autocorrelation functions for suitably defined parameters of the algorithm is studied. Section~\ref{sec:summary} contains a discussion of the results.  

\section{Model}\label{2}
	Consider a collection of particles which are labeled by their masses. Given a configuration, the system evolves in time through mass-conserving binary aggregation: 
	\begin{equation}
	A_i + A_j  \xrightarrow{K(i,j)} A_{i+j},
	\end{equation}
	where $A_k$ denotes a particle of mass $k$, and the collision kernel $K(i,j)$ is the rate at which two particles of masses $i$ and $j$ aggregate. In an infinitesimal time $dt$, the probability of collision of two particles having masses $i$ and $j$ is given by $K(i,j)dt$.
	Since each aggregation event reduces the number of particles, $N(t)$, by $1$, $N(t)$ decreases monotonically with time. Initially, there are $N(0)=M$ particles with equal mass $m_0$. We set $m_0=1$, so that all masses are measured in units of $m_0$.
	
We are interested in the probability distribution $P(M,N,t)$, defined as the probability of $t$ being the minimum time at which exactly $N$ particles are remaining, or equivalently the probability that the $(M-N)^{th}$ collision occurs at time $t$, given that there are $M$ particles of mass $1$ initially. Here, we consider $t$ as the random variable with $\int_0^\infty dt P(M,N,t)=1$. Also, we would like to know what the most probable trajectory is for a given $M,N,t$.
	
When $t$ is the typical time for given $M$ and $N$, then we expect that the most probable trajectory is described by the Smoluchowski equation:
	\begin{equation}
	\frac{d N_i(t)}{dt}=\frac{1}{2}\sum_{m_1}\sum_{m_2}K(m_1,m_2)N_{m_1}N_{m_2}\delta(m_1+m_2-i)-N_i\sum_{m_1}K(i,m_1)N_{m_1},\label{smol}
	\end{equation}
where $N_i(t)$ is the number of particles of mass $i$ at time $t$. This equation is solvable for the typical trajectory for only few collision kernels: constant, sum and product~~\cite{leyvraz2003scaling,aldous1999deterministic,krapivsky2010kinetic,handbook}. We note that the Smoluchowski equation ignores correlations among the particles, and also does not give any information about atypical times, the focus of this paper.

	\section{Results}\label{results}
	\subsection{Monte Carlo Algorithm \label{3}}
\begin{figure}
		\centering
		\includegraphics[width=\columnwidth]{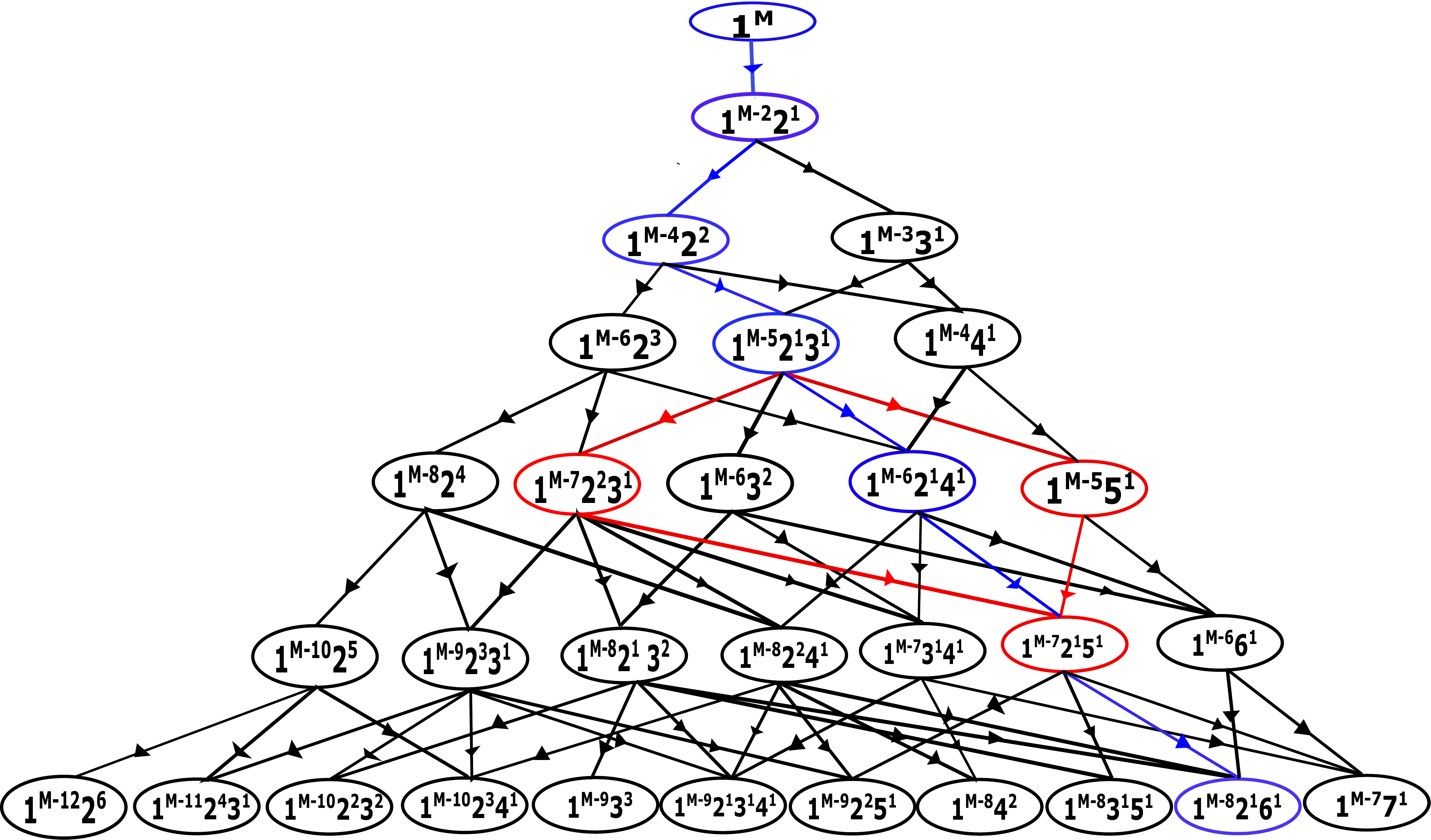}
		\caption{All possible configurations and trajectories for $6$ collisions. The configurations after each collision are shown inside the bubbles. The bubbles at a certain level are arranged from left to right according to the order relation described in text (see Sec.~\ref{ergodicity}). For the trajectory shown in blue, the  red lines denote possible alternate paths that alter only the 4-th configuration.}
		\label{fig:collisions2}
	\end{figure}

	We now describe a Monte Carlo algorithm to numerically determine $P(M,N,t)$ for any given aggregation kernel. This includes times which are atypical for a given $M,N$, and hence are dominated by rare events. A trajectory that contributes to $P(M,N,t)$ consists of $C=M-N$ collisions. As $C$ increases, the number of trajectories increases rapidly.  Figure (\ref{fig:collisions2}) shows all the possible configurations for $6$ collisions. Any path from the top row to the bottom row along the directed edges constitutes a trajectory.

	To compute $P(M,N,t)$ for atypical times $t$, we use a method known as importance sampling~\cite{denny2001introduction}. The simulations are performed at constant $M$ and $N$, and $t$ is considered as the random variable. In addition to weights arising from the aggregation kernel, each trajectory is weighted by $e^{wt},$ where $w$ is a biasing parameter which can be positive or negative. Thus, the biased distribution is 
	\begin{equation}
	P_w(M,N,t)=\frac{1}{Z}P(M,N,t)e^{wt},\label{eq:unbias}
	\end{equation}
	where $Z$ is a normalizing factor. Positive $w$ biases the system towards larger times and negative $w$ towards smaller times, resulting in robust sampling of atypical trajectories. We first determine $P(M,N,t)$ without bias, \textit{i.e.,} for $w=0$. Then, we obtain $P_w(M,N,t)$ for $w\neq 0$ and unbias the distribution using Eq.~(\ref{eq:unbias}), i.e., multiplying by $e^{-wt}$. To combine the data obtained from different choices of $w$, we proceed as follows. The base normalized distribution is the unbiased distribution of $P(M,N,t)$ obtained for $w=0$. The values of $w$ are chosen such that between two successive choices of $w$, there is some overlap in the sampled times. The biased distribution is glued on by minimizing the error in the data for the overlapping times.
	
	The probability distribution $P(M,N,t)$ is a sum over the probabilities of each trajectory with $C$ collisions. A trajectory is characterized both by the sequence of collisions as well as the waiting times between consecutive collisions. In the Monte Carlo algorithm, we introduce local modifications to the trajectory by changing both of the above, as described below.
	
	To characterize a trajectory, we introduce the following notation. We will refer to the configuration after the $i$-th collision as the $i$-th configuration. Its mass distribution, the number of particles of mass $m$,  will be denoted by $N_{i}(m)$. Note that it suffices to give either the sequence of collisions or the configurations to specify the trajectory. The waiting time between the $i$-th and the $(i+1)$-th collisions, or equivalently the waiting time for the $i$-th configuration, will be denoted by $\Delta t_i$. Also, $(m_i,m_i^\prime)$ will refer to the pair of masses aggregating in the $i$-th collision.
	
	At each micro-step, a configuration is chosen uniformly at random, say the $i$-th configuration. With probability $p$, the waiting time, $\Delta t_i$, associated with the $i$-th configuration, is modified, keeping all the configurations fixed. With probability $(1-p)$, the $i$-th configuration is modified, keeping all other configurations as well as all waiting times fixed. We will treat $p$ as a parameter of the algorithm.
	
	We first describe the change in waiting times. Let the current waiting time for the $i$-th configuration be denoted by $\Delta t_i^{old}$. A new waiting time $\Delta t_i^{new}$ is drawn from an exponential distribution~\cite{gillespie1975exact}
	\begin{equation}
	P(\Delta t_i)=\mathcal{R}_i e^{-\mathcal{R}_i \Delta t_i},\label{wt} 
	\end{equation}
	where $\mathcal{R}_i$ is the total rate of collision of the $i$-th configuration. 
	In terms of the collision kernel, \begin{equation}\mathcal{R}_i=\sum_{m_1=1}^M\sum_{m_2\geq m_1}^M K(m_1,m_2)\mathcal{C}_i(m_1,m_2),\label{rate}\end{equation} where 
	\begin{equation}\mathcal{C}_i(m_1,m_2)=N_i({m_1})N_i({m_2})-\frac{N_i({m_1})\left[N_i({m_2})+1\right]}{2}\delta_{m_1,m_2},\label{comb}\end{equation} is the combinatorial factor associated with the number of ways of choosing particles of masses $m_1$ and $m_2$.
	For ensuring detailed balance, we first note that when the waiting times are changed the sequence of collisions remains the same for both the old and new trajectories. Since the waiting times are biased with weight $e^{-wt}$, it is easy to see that detailed balance is satisfied if the new waiting time $\Delta t_i^{new}$ is accepted with a probability $\min[1, e^{w(\Delta t_i^{new}-\Delta t_i^{old})}]$. 
	
Second, we describe the moves to modify the trajectory through changes in the configurations. 
	There are multiple ways of choosing a different pair of successive collisions such that only the $i$-th configuration is changed. An example of possible options is shown in Fig.~\ref{fig:collisions2}. Consider the trajectory shown in blue.  To change the $4$-th configuration, keeping other configurations fixed, the paths that are marked in red are also allowed, but each with different weights. We now formulate the general rules to obtain the set of collisions which will preserve all configurations except the $i$-th configuration.
	
	If $1<i<C$, then the $i$-th and $(i+1)$-th collisions have to be modified, while if $i=C$, only the $C$-th collision has to be modified. We first discuss the case $1<i<C$. For convenience of notation, let the pair of successive collisions be denoted as $(m_1,m_2)$,$(m_3,m_4)$ respectively. The most obvious way that the collisions can be modified is to reverse the sequence of collisions such that the collision $(m_3,m_4)$ occurs first, and then $(m_1,m_2)$ occurs, provided the masses  $m_3$ and $m_4$ exist independent of the $(m_1,m_2)$ collisions. The collisions can also occur such that the product from the $i^{th}$ collision, \textit{i.e.,} $m_1+m_2$,  is one of the colliding masses of the next collision, say $m_3$. This possibility leads to the classification of the pair of collisions into three types. A pair of collisions where $m_1+m_2\neq m_3$ or $m_4$ can undergo only reversal of the sequence of collisions. This type of collision will be denoted as $\alpha$. A pair of collisions where $m_1+m_2=m_3$ falls into two types, $\beta$ and $\gamma$. All the three types are described below. The rules pertaining to all three types are given in Table~\ref{moves}.
	
		Type $\alpha$:  $m_{1}+m_2\neq m_3$ or $m_4$. Here the only possibility is to reverse the sequence of collisions and thus there are only two pairs of collisions to choose from.
		
			Type $\beta$ : $m_{1}+m_{2}= m_{3},$ and there is at least one particle of mass $m_{3}$ in the $(i-1)$-th configuration. In this case, there are $6$ possible pairs of collisions to choose from.
			
		Type $\gamma$: $m_{1}+m_{2}= m_{3},$ but there are no particles of type $m_{3}$ in the $(i-1)$-th configuration. Compared to type $\beta$, the pair of reversed collisions $(m_{3},m_{4}),(m_1,m_2)$ would not occur. Thus, there are $5$ possible pairs of collisions to choose from.
	
	Each of the possibilities in Table~\ref{moves} occurs with weight,
	\begin{align}
	\begin{aligned}
	W(m_{1},m_{2};m_{3},m_{4})= &K(m_1,m_2)\mathcal{C}_{i-1}(m_1,m_2)\mathcal{R}_{i-1}e^{-\mathcal{R}_{i-1}\Delta t_{i-1}} \\&K(m_{3},m_{4})\mathcal{C}_{i}(m_{3},m_{4})\mathcal{R}_{i}e^{-\mathcal{R}_{i}\Delta t_{i}}.
	\end{aligned}
	\end{align}
	If $i=C$, \textit{i.e.,} the $C$-th configuration is chosen, then any two masses from the $(C-1)$-th configuration may aggregate. For the final collision, the weight of choosing a pair is 
	\begin{equation}
	W(m_{C},m^\prime_{C})= K(m_{C},m^\prime_{C})\mathcal{C}_{C-1}(m_{C},m^\prime_{C})\mathcal{R}_{C-1}e^{-\mathcal{R}_{C-1}\Delta t_{C-1}}. 
	\end{equation}
	\begin{table}
		\caption{Given a pair of successive collisions $(m_1,m_2)$ and $(m_3,m_4)$, all allowed alternate pairs of collisions that alter only the intermediate configuration $i$ are tabulated. If the product of the first collision takes part in the second, we denote the product as $m_3$ without loss of generality.}
		\begin{tabular}{|c|c|l|l|} \hline			Type & Description&$i$-th collision & $(i+1)$-th collision\\ \hline
			$\alpha$  & $m_{1}+m_{2}\neq m_{3}$\label{gamma} &
			\begin{tabular}{l} $(m_{1},m_{2})$ \\ $(m_{3},m_{4})$
			\end{tabular} &
			\begin{tabular}{l} $(m_{3},m_{4})$ \\ $(m_{1},m_{2})$
			\end{tabular}
			\\ \hline
				 $\beta$ & \begin{tabular}{l} $ m_{1}+m_{2}= m_{3}$,\\ \\ $N_{i}(m_{3})>0$ \end{tabular} \label{alpha} &
			\begin{tabular}{l} 
				$(m_{1},m_{2})$ \\ $(m_{3},m_{4})$ \\ $(m_{1},m_{4})$ \\ $(m_{1}+m_{4},m_{2})$ \\$(m_{2},m_{4})$ \\ $(m_{2}+m_{4},m_{1})$
			\end{tabular} &
			\begin{tabular}{l} $(m_{3},m_{4})$ \\ $(m_{1},m_{2})$ \\ $(m_{1}+m_{4},m_{2})$ \\$(m_{1},m_{4})$  \\ $(m_{2}+m_{4},m_{1})$ \\ $(m_{2},m_{4})$
			\end{tabular} \\ \hline
			$\gamma$ & \begin{tabular}{l}$ m_{1}+m_{2}= m_{3}$,\\ \\$ N_{i}(m_{3})=0$ \end{tabular}\label{beta} &
			\begin{tabular}{l} $(m_{1},m_{2})$  \\ $(m_{1},m_{4})$ \\ $(m_{1}+m_{4},m_{2})$ \\$(m_{2},m_{4})$ \\ $(m_{2}+m_{4},m_{1})$
			\end{tabular} &
			\begin{tabular}{l} $(m_{3},m_{4})$ \\ $(m_{1}+m_{4},m_{2})$ \\$(m_{1},m_{4})$  \\ $(m_{2}+m_{4},m_{1})$ \\ $(m_{2},m_{4})$
			\end{tabular} \\ \hline
		
		\end{tabular}
		\label{moves}
	\end{table} 
	
	From all the allowed possibilities, we choose a particular configuration with probability proportional to its weight, thus making the choice rejection-free. A Monte Carlo move consists of
	$2C$ micro-steps.
	
	The algorithm obeys detailed balance. Once the set of new configurations is determined, based on the current configuration, the probability of choosing a particular configuration is only proportional to its weight, and independent of the current configuration. Hence, the configurational moves satisfy detailed balance trivially. The assignment of waiting times follows the usual Metropolis rule and hence satisfies detailed balance. 
	
\begin{figure}
		\centering
		\includegraphics[width=0.8\columnwidth]{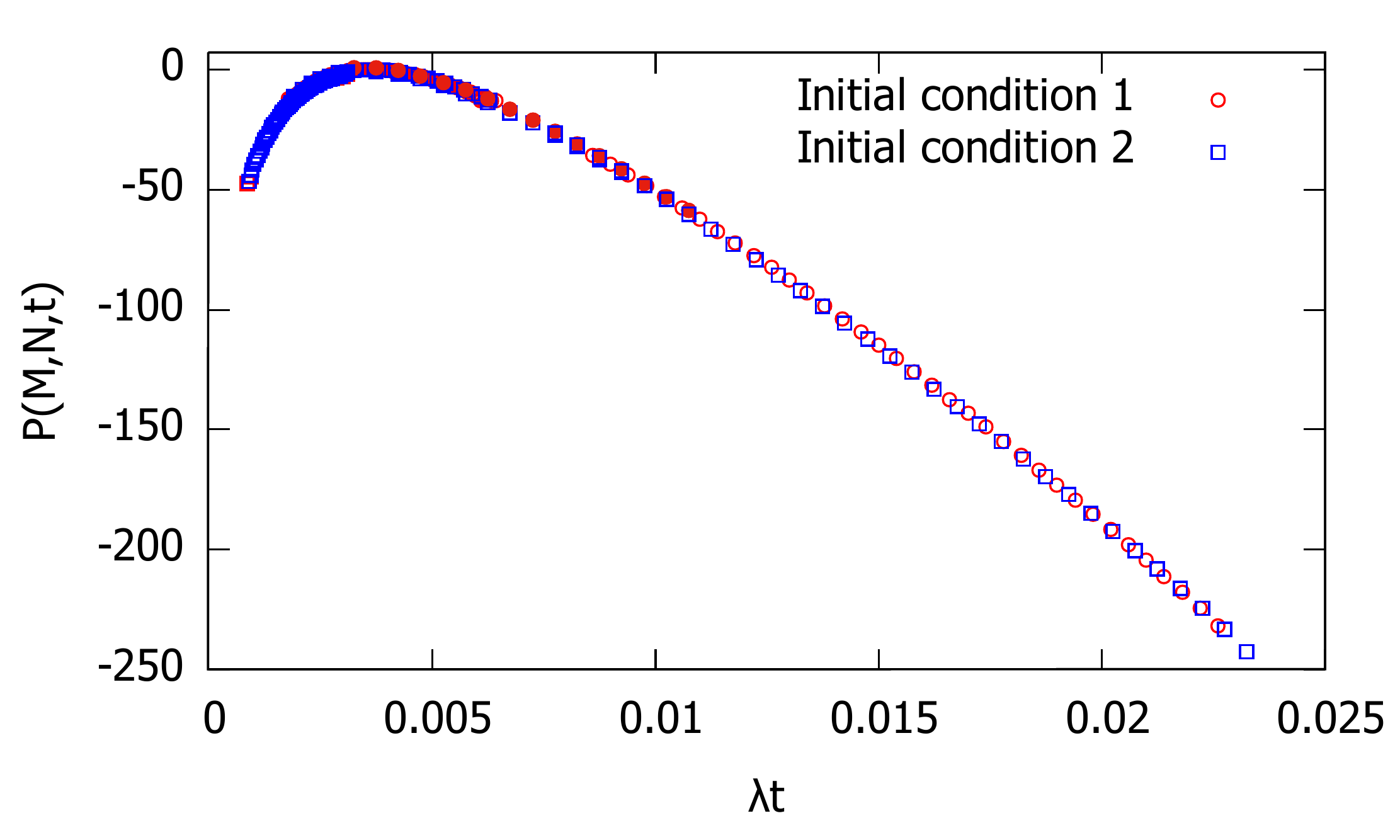}
		\caption{ Comparison of $P(M,N,t)$ for $M=240$, $N=168$, obtained for two different initial conditions. Initial condition 1 corresponds to the initial trajectory consisting of $C=M-N$ random collisions. Initial condition 2 corresponds to the trajectory $(\textbf{1})^{M-i}(\textbf{i})^1$, formed by the collision of a particle of mass $1$ with a particle of mass $M-i$. The data are for the constant kernel.}
		\label{fig:init}
	\end{figure}

	The initial configuration is chosen  by colliding a randomly chosen pair of particles at each collision. The initial waiting times are drawn from the exponential distribution Eq.~(\ref{wt}).  
To confirm convergence, we check that the results do not depend on the initial trajectory, by choosing other initial trajectories such as $(\textbf{1})^{M-i}(\textbf{i})^1$. As an example, in Fig.~\ref{fig:init} we compare $P(M,N,t)$ for the constant kernel,  obtained for the two initial conditions discussed above. The data are indistinguishable from each other, confirming equilibration.
	
	\subsection{Ergodicity of the Monte Carlo Algorithm}\label{ergodicity}
	The Monte Carlo algorithm modifies trajectories using local moves which are reversible and obey detailed balance. We now show that the algorithm is ergodic, $\textit{i.e.,}$ it allows all trajectories to be accessed. To do so, it is enough to prove that an arbitrary trajectory, $A$, can transform to a standard trajectory, $S$, through a given protocol. Then, to transform $A$ to any given trajectory $B$, we follow the protocol from $A$ to $S$ and reverse the moves from $B$ to $S$.
	
	We choose the standard trajectory $S$ to be one where after every collision, only one mass different from $1$ is allowed at all times. That is, after $i$ collisions, the configuration is $(\textbf{1})^{M-i-1}(\textbf{i+1})^1$. In this trajectory, at each collision, the largest mass collides with a particle of mass $1$.
	
	To describe the protocol of transforming an arbitrary trajectory to $S$, it is convenient to introduce an ordering among configurations that have undergone the same number of collisions. We will say that $(\textbf{1})^{N_1}(\textbf{2})^{N_2}\dots<(\textbf{1})^{N_1^{\prime}}(\textbf{2})^{N^{\prime}_2}\dots$ if $N_1=N^{\prime}_1, N_2=N^{\prime}_2,\dots,N_{k-1}=N^\prime_{k-1,} N_k<N^{\prime}_k$ , where $k$ is the smallest mass for which $N_k\neq N^{\prime}_k$. The configurations are then arranged in increasing order, as shown in Fig. (\ref{fig:collisions2}). In this representation, the standard trajectory $S$ is the rightmost trajectory.
	
\begin{figure}
		\centering
		\includegraphics[width=\columnwidth]{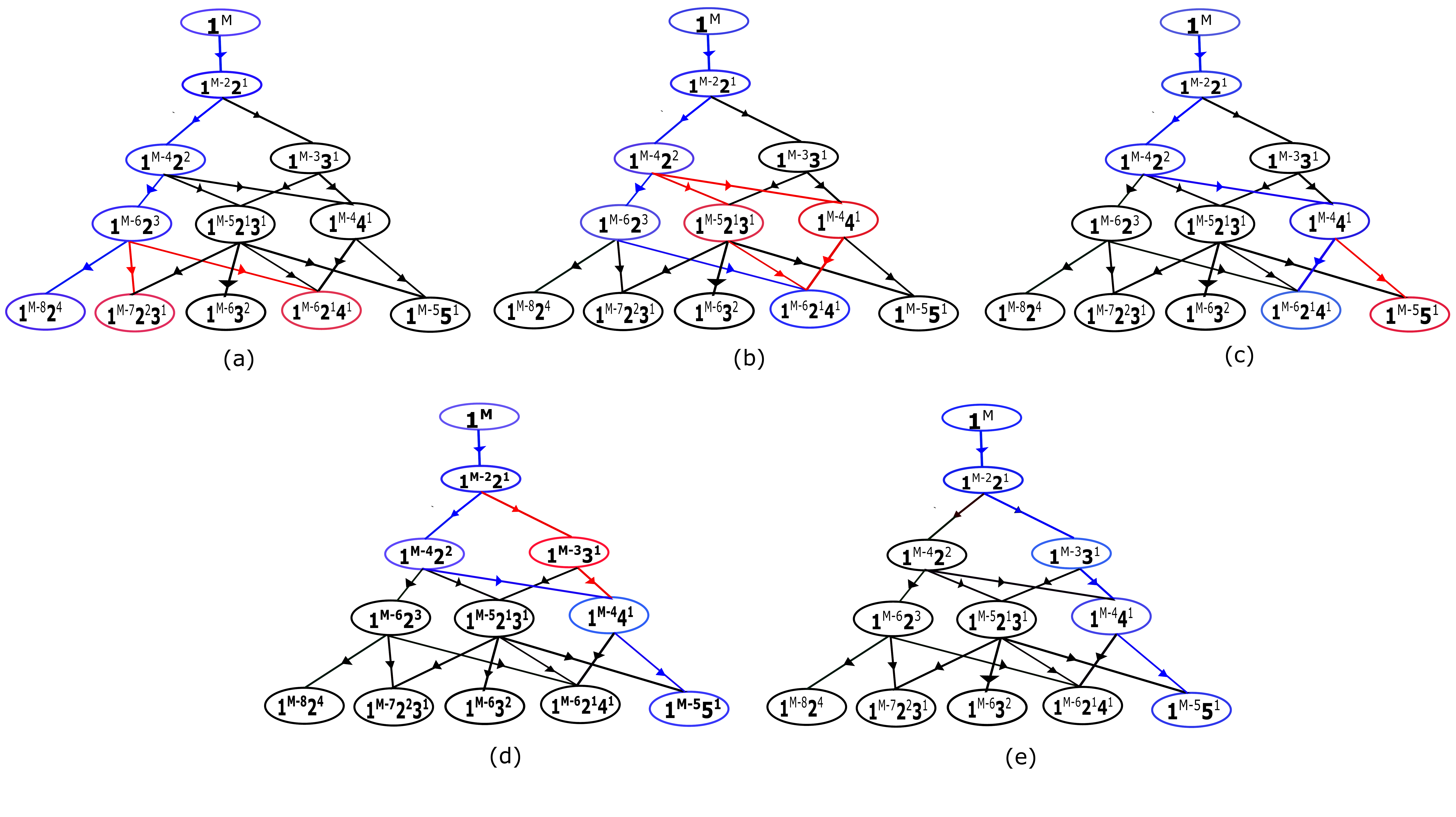}
		\caption{The transformation from the leftmost trajectory [blue path in (a)] where all the configurations result from the collision $(1,1)$,  to the rightmost trajectory, $S$, using the protocol described in the text. The current path is depicted in blue, and the possible transformations in the next collision, as prescribed by the protocol, are denoted in red.}
		\label{traj}
	\end{figure}

	Consider any arbitrary trajectory $A$. The following transformations are applied till no more transformation is possible :
	\begin{itemize}
		\item The lower most edge is moved to the rightmost allowed node.
		\item For the bottom most configuration that can be modified such that the trajectory moves rightward, we choose the rightmost path.
	\end{itemize}
	We  give an example of the above protocol for a trajectory with $4$ collisions. Consider the leftmost trajectory shown in blue, in Fig. \ref{traj}(a) where the configuration after $i$ collisions is $(\textbf{1})^{M-2i}(\textbf{2})^i$. 
	The protocol transforms the trajectory as follows. The lowest-most edge has two other valid choices as shown in red in Fig.~\ref{traj}(a). We choose the rightmost of these to obtain the blue trajectory in Fig.~\ref{traj}(b). The third configuration now can be moved rightwards along the paths shown in red in Fig.~\ref{traj}(b). We choose the right-most configuration to obtain the blue trajectory in Fig.~\ref{traj}(c). The bottom-most edge is now moved to the edge shown in red in Fig.~\ref{traj}(c) to obtain the blue trajectory in Fig.~\ref{traj}(d). Finally, the second configuration is moved to the right along the red path shown in Fig.~\ref{traj}(d) to obtain the standard trajectory shown in blue in Fig.~\ref{traj}(e).

	We now show that the protocol transforms an arbitrary trajectory A to the standard trajectory $S$. Suppose, on application of the protocol, $A$ is transformed to $S'$. We will now show that $S^\prime=S$. 
	Let the sequence of collisions in $S^{\prime}$ be denoted by $(m_i,m^\prime_i)$, $\textit{i.e.,}$ in the $i$-th collision, masses $m_i$ and $m^{\prime}_i$ aggregates. We derive the constraints that two consecutive collisions in $S^\prime$, $\left[(m_i,m^{\prime}_i), (m_{i+1},m^\prime_{i+1})\right]$ should obey.
	A pair of collisions falls under one of the three types,  $\alpha,\beta$ and $\gamma$ as described in Sec.~\ref{3}.

The argument is based on the following observation. Suppose, given a configuration, we consider two possible collisions: $(m_1, m_2)$ or $(m_3, m_4)$. If the trajectory due to $(m_3, m_4)$ colliding is to the right of the trajectory due to $(m_1, m_2)$ colliding, then  based on the order relation, it is easy to see that $\min(m_1, m_2)\leq \min(m_3, m_4)$.
	
For collisions of the type $\alpha$, there are two possibilities (see Table~\ref{moves}) for the sequence of collision. The reversed sequence of collisions $(m_{i+1},m^\prime_{i+1})$,  $(m_{i},m^\prime_{i})$ would lead to a trajectory to the right of $S^\prime$ if $\min(m_{i+1},m^\prime_{i+1})>\min(m_i,m^\prime_i)$. Since we cannot have a trajectory that is to the right of $S^\prime$,  we obtain the condition, based on the argument in the previous paragraph,
	$\min(m_{i+1},m^\prime_{i+1})\leq\min(m_{i},m^\prime_i).$	
	
	Now consider collisions of types $\beta$ and $\gamma$. The pair of collisions are $(m_i,m^\prime_i)$ and $(m_i+m^\prime_i,m^\prime_{i+1})$. Suppose $m^\prime_{i+1}>\min(m_i,m^\prime_i)$. Then the pair of collisions $(m^\prime_{i+1},\max(m_i,m^\prime_i))$, $(\min(m_i,m^\prime_i),m^\prime_{i+1}+\max(m_i,m^\prime_i))$ creates a trajectory to the right of $S^\prime$ that is allowed by the protocol. But since $S^\prime$ is the rightmost trajectory, there is a contradiction and hence
	\begin{equation}
	 m^\prime_{i+1}=\min(m_{i+1},m^\prime_{i+1})\leq \min(m_i,m^\prime_i),\label{condition1}
	 \end{equation} 	that holds for all collision types $\alpha,\beta$ and $\gamma$.
	 
	 The first collision is $(1,1)$. To satisfy the condition in Eq. (\ref{condition1}), it is clear that at least one of the colliding masses in the second collision should be $1$, as the minimum possible mass is $1$. It follows that in order to satisfy the condition in Eq. \ref{condition1} for every sequence of consecutive collisions in the trajectory, at least one of the colliding masses in all the subsequent collisions should be $1$.   
	
	Now consider the $C$-th collision. For the rightmost trajectory, the two largest masses have to be collided. But we have already shown that one of the masses should be $1$, \textit{i.e.}, the second largest mass is $1$.  This implies that the $C$-th configuration is $\textbf{1}^{M-1}\textbf{C}^1$. Using the property that mass $1$ is used in each step, it follows that the $i$-th configuration is $\textbf{1}^{M-i}i^1$, which is the standard configuration. This implies that $S^\prime=S$, and hence proves that the algorithm is ergodic. 

	\subsection{Large Deviation Function}\label{ldev}
	To show the efficacy of the algorithm, we compare the numerical results with the exact solution of the model of constant kernel where collision rates are independent of the masses.  The collision kernel $K(m_1,m_2)=\lambda$.
	\subsubsection{Exact result for constant kernel}
	When the collision rates are independent of masses, $P(M,N,t)$ can be analytically computed.
	After $i$ collisions, $M-i$ particles remain, and the total rate of collision is given by  \begin{equation}
	\mathcal{R}_i=\frac{\lambda(M-i)(M-i-1)}{2}.\label{rates}\end{equation} Using the exponential time distribution Eq.(\ref{wt}), 
	\begin{align}
	\begin{aligned}
	P(M,N,t)&=\int_{0}^{\infty}d\Delta t_0\int_{0}^{\infty}d\Delta t_1...\int_{t=0}^{\infty}d\Delta t_{C-1}  ~~\mathcal{R}_0 e^{-\mathcal{R}_0 \Delta t_0} \\&\mathcal{R}_1e^{-\mathcal{R}_1 \Delta t_2}\dots\mathcal{R}_{C-1}e^{-\mathcal{R}_{C-1}\Delta t_{C-1}}\delta\left(\sum_{i=0}^{C-1}\Delta t_i-t\right).
	\end{aligned}
	\end{align}
	The $\delta$-function constrains the sum of waiting times to the total time $t$. The Laplace transform of $\widetilde{P}(M,N,s)$, defined as 
	\begin{equation}
	\widetilde{P}(M,N,s)=\int_{0}^{\infty}dt e^{-st}P(M,N,t),
	\end{equation}
	is then
	\begin{equation}
	\widetilde{P}(M,N,s)=\prod_{i=0}^{C-1}\frac{\mathcal{R}_i}{\mathcal{R}_i+s}.
	\end{equation}
	Doing the inverse Laplace transform, we obtain
	\begin{equation}	
	P(M,N,t)=\left(\prod_{k=0}^{C-1}\mathcal{R}_k\right)\sum_{i=0}^{C-1}e^{-\mathcal{R}_i t} \prod_{j\neq i,{j=0}}^{C-1}\frac{1}{\mathcal{R}_j-\mathcal
		{R}_i}\label{soln}.
	\end{equation}
	
\begin{figure}
		\centering
		\includegraphics[width=0.85\columnwidth]{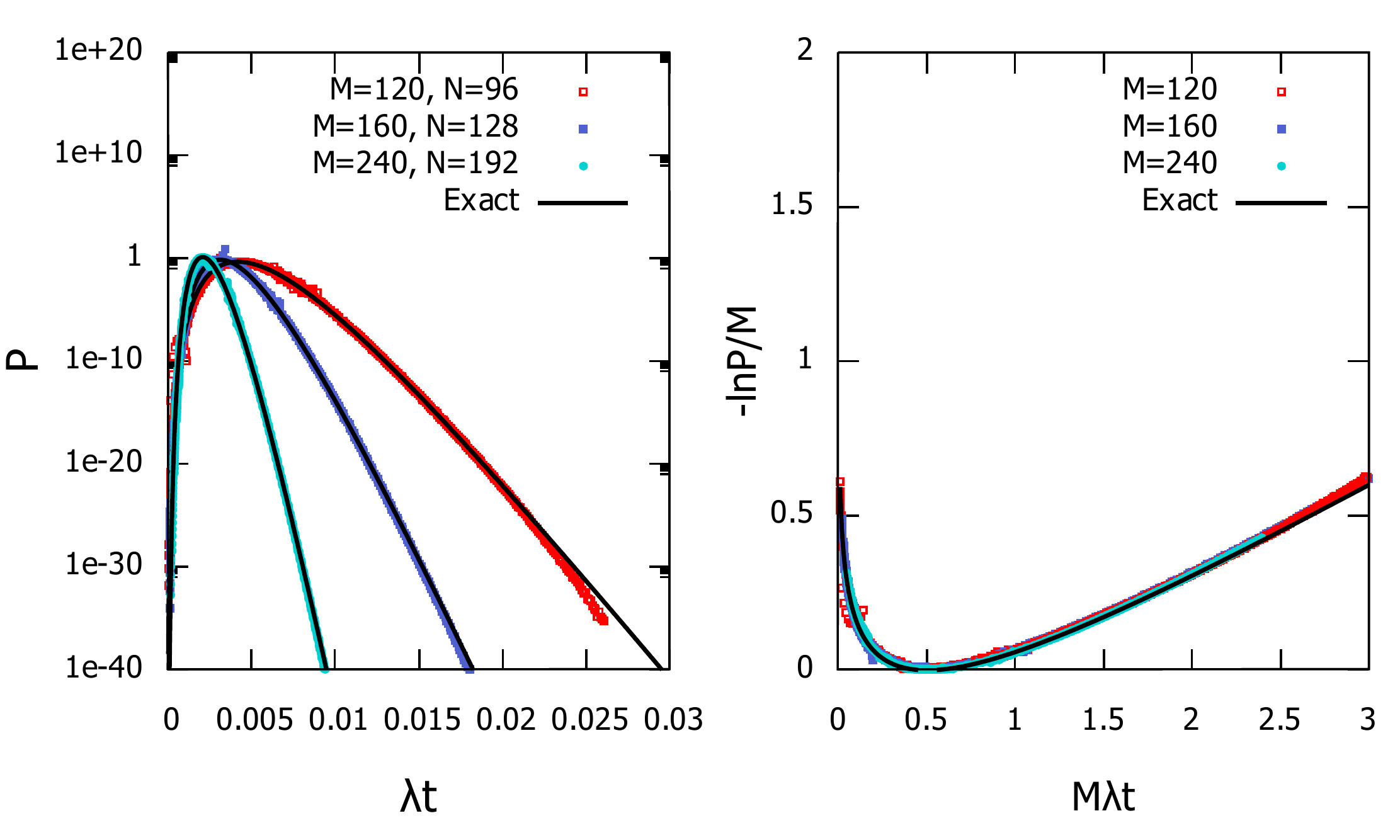}
		\caption{(a) $P(M,N,t)$ for the constant kernel for different $M$, keeping $\Phi=N/M=0.8$ fixed are compared with the exact solution, Eq.~(13). (b) The data in (a) for different $M$ collapse onto one curve when scaled as in Eq.~(14), to give the rate function.}
		\label{fig:phi0}
	\end{figure}

	We compare the results from the Monte Carlo simulations for the constant kernel with the exact results. Plotting the unbiased  $P(M,N,t)$ (with $w=0$) as the reference, $P(M,N,t)$ obtained from non-zero values of $w$ are merged with the reference distribution by appropriate normalization. In Fig.~\ref{fig:phi0}(a), the results for $P(M,N,t)$ from Monte Carlo simulations are compared with the exact solution for a fixed $\Phi=N/M=0.8$ and $M=120, 160, 240$. It is clear that the data are in good agreement with the exact results, thus providing a benchmark for correctness. Also, we are able to measure very low probabilities, of the order of $10^{-35}$, and even lower, at times much larger and much smaller than the typical time.
	
 For large $M$, $P(M,N,t)$ for different $M,N,t$ collapse onto one curve when scaled as in
	\begin{equation}
	-\ln P(M,N,t)= M f\left(\frac{N}{M}, M\lambda t\right),~~~~ M, N,t^{-1} \to \infty,\label{ld}
	\end{equation}
	as shown in Figure~\ref{fig:phi0}(b). We then identify $M$ with the rate and $f$ with the large deviation function~\cite{touchette2009large}. $-\ln P$ has a minimum value of zero. We will identify the corresponding value of time as the typical time, $t_{typ}$ for $M-N$ collisions, \textit{i.e.,} $f(\Phi, M\lambda t_{typ})=0$.

\begin{figure}
		\centering
		\includegraphics[width=0.85\columnwidth]{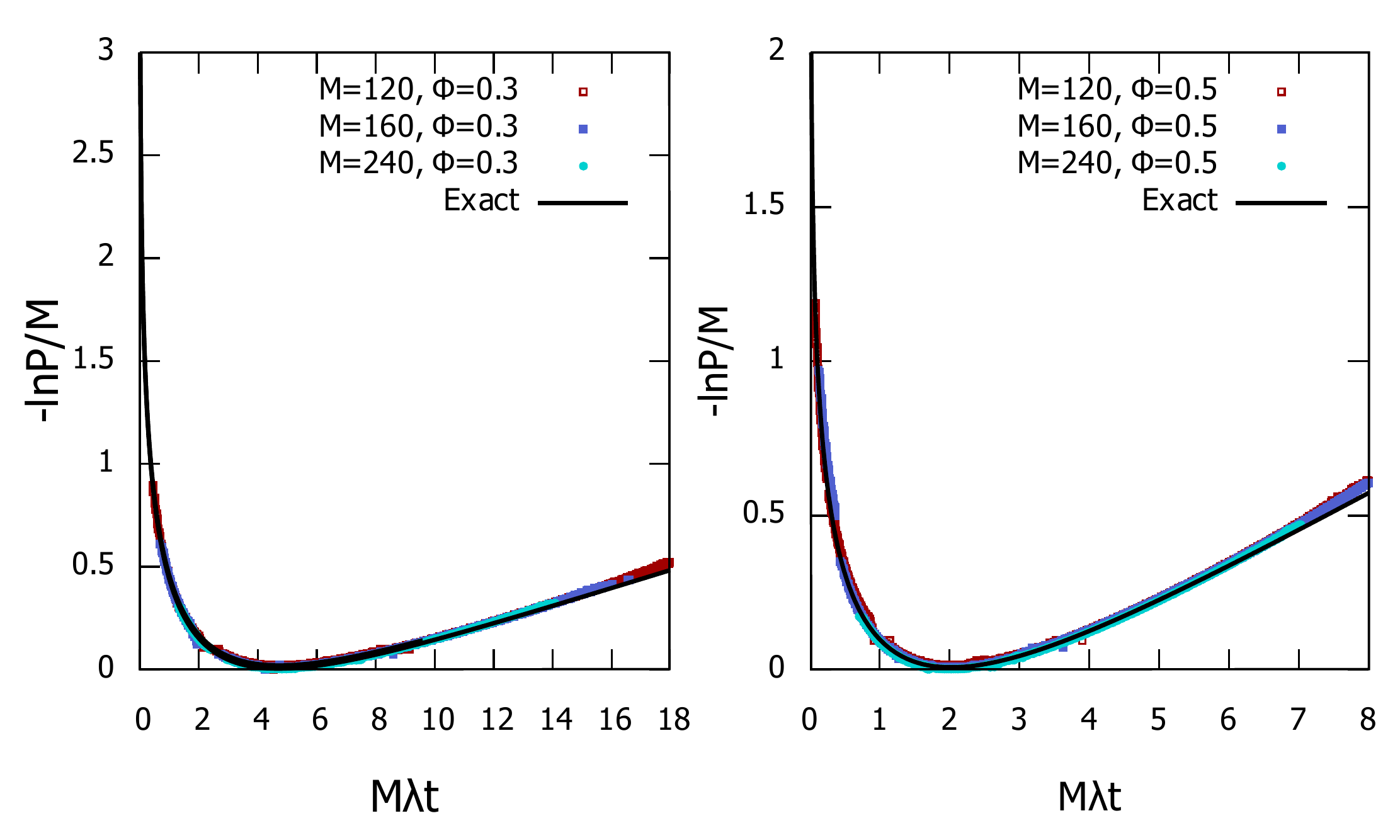}
		\caption{(a) $P(M,N,t)$ for the constant kernel for different $M$,  and keeping $\Phi=N/M=0.3$  fixed collapse onto one curve when scaled as in Eq.~(\ref{ld}), to give the rate function. (b) $P(M,N,t)$ for the constant kernel for different $M$,  and keeping $\Phi=N/M=0.5$  fixed collapse onto one curve when scaled as in Eq.~(\ref{ld}), to give the rate function. }
		\label{fig:ratefn-scaled}
	\end{figure}

		To show that the algorithm works for the full range of $\Phi$, we compare the results from simulations of the constant kernel with the exact results for $\Phi=0.3,0.5$ in Fig.~\ref{fig:ratefn-scaled}. Excellent agreement is seen.

	We note that there is an upper bound for the value of the bias  $w$. To see this, we observe that $P(M,N,t)$ in Eq.~(13) is a sum over $(C+1)$ terms, each one of which decreases exponentially with $t$ as $e^{-R_i t}$. Thus, for large $t$, the term with the smallest $R_i$ will dominate. Since the smallest rate is $R_C$, we expect that
	\begin{equation} P(M,N,t)\approx \mathcal{R}_{C-1}e^{-\mathcal{R}_{C-1} t} \prod_{{j=0}}^{C-2}\frac{\mathcal{R}_j}{\mathcal{R}_j-\mathcal
		{R}_C-1},~~ t\to \infty.\end{equation}
 	This implies that a bias $w>\frac{1}{2}N(N+1)$ cannot be applied since the biased distribution $P_w(M,N,t)$ would diverge, making it not normalisable. For small times, there is no such cutoff for the bias.
	\begin{figure}
		\centering
		\includegraphics[width=0.8\columnwidth]{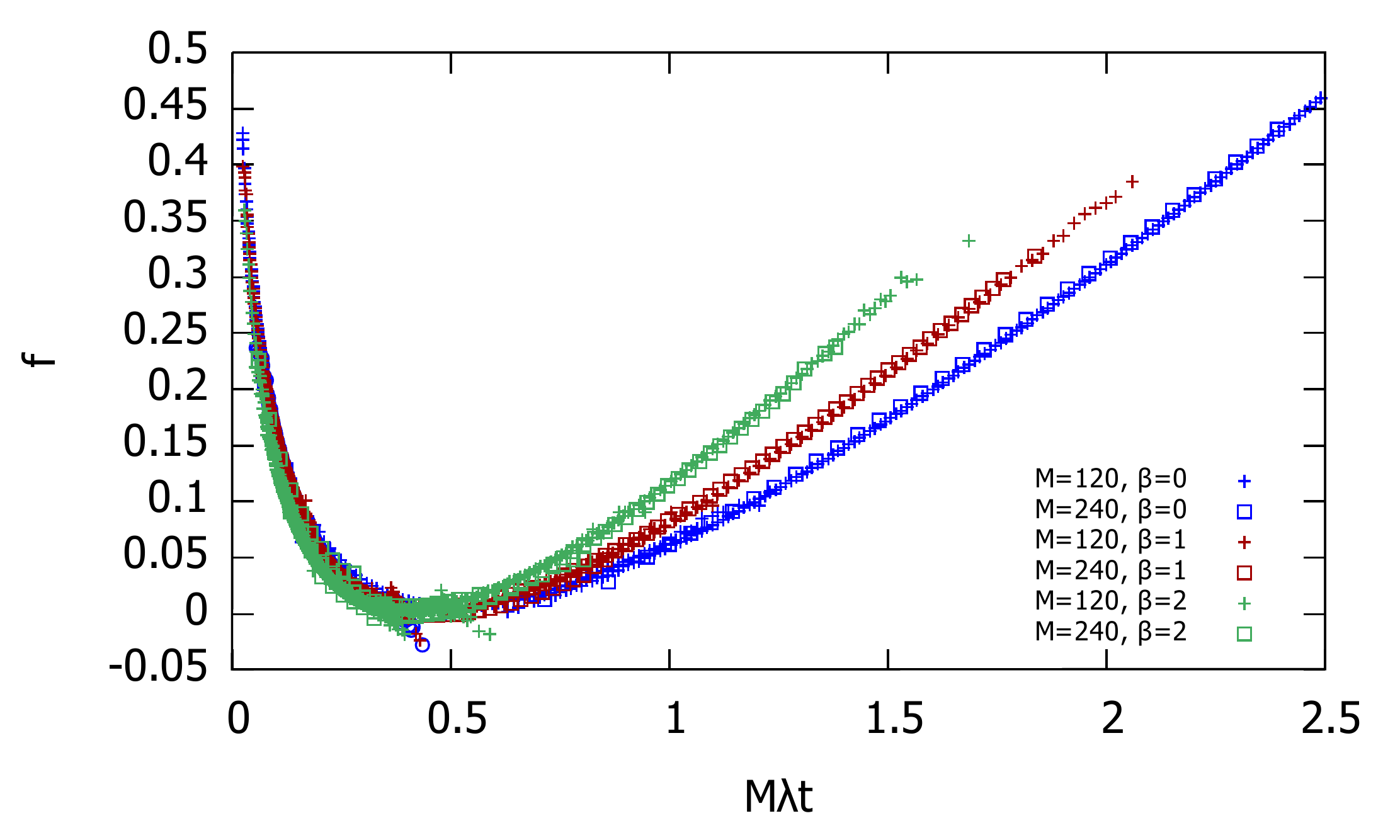}
		\caption{The large deviation rate function $f(\Phi, M\lambda t)$ [see Eq.~(\ref{ld})] for fixed $\Phi=N/M=0.8$ and different kernels, $K(m_1,m_2)=(m_1m_2)^{\frac{\beta}{2}}$ where $\beta=0,1,2$. The data are for $M=120$ and $M=240$.}
		\label{fig:otherkernels}
	\end{figure}
	
	\begin{figure}
		\centering
		\includegraphics[width=0.85\columnwidth]{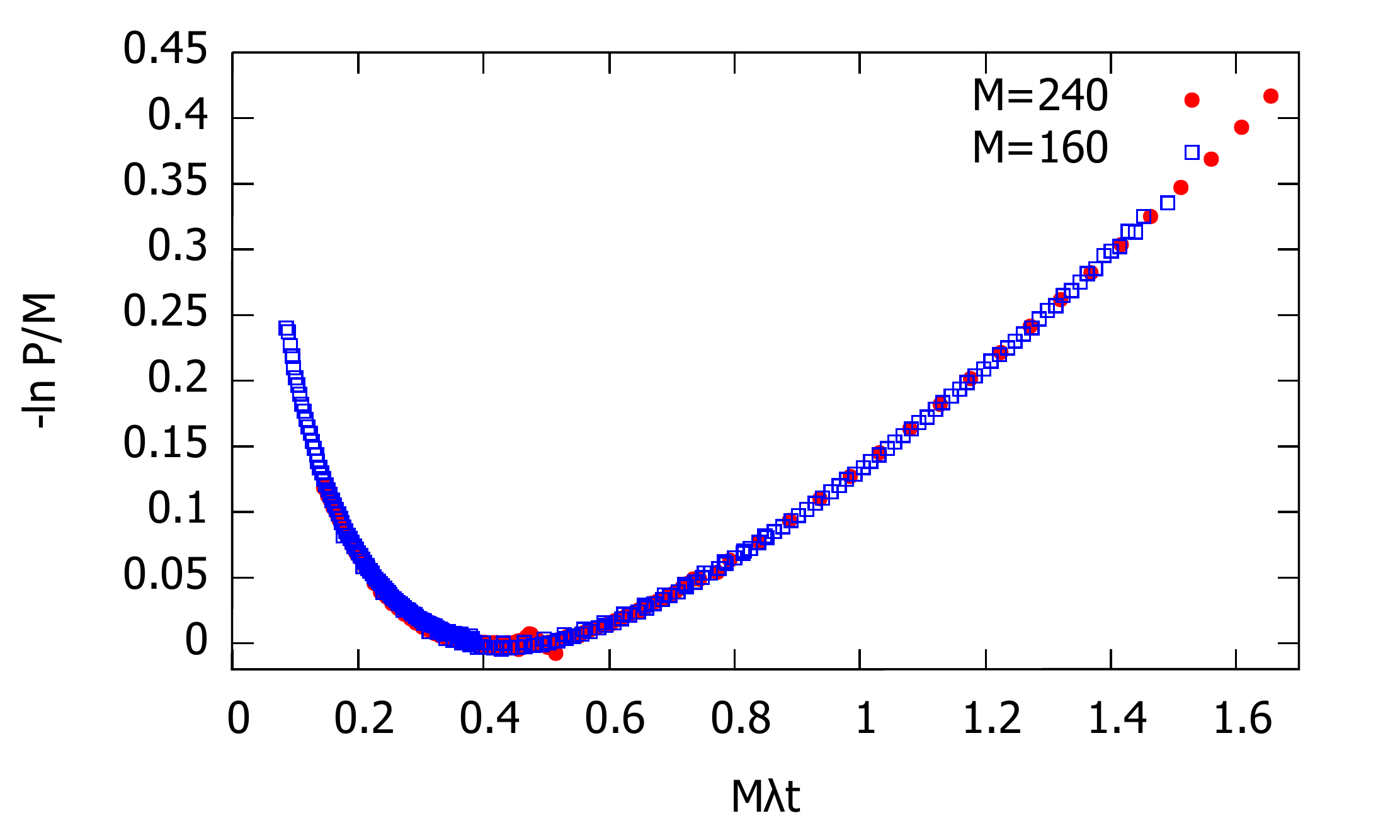}
		\caption{The large deviation rate function $f(\Phi, M\lambda t)$ [see Eq.~(\ref{ld})] for the Brownian kernel, $K(m_1,m_2)=(\frac{m_1}{m_2})^{1/3}+(\frac{m_2}{m_1})^{1/3}+2$, for $\Phi=N/M=0.7$. The data are for $M=160$ and $M=240$.}
		\label{fig:brownian}
	\end{figure}
	
	The large deviation functions of kernels other than the constant kernel can also be obtained using the algorithm.  Depending on the form of the collision kernel $K(m_1,m_2)$, a phenomenon known as gelation occurs in aggregating systems, where there is a non-trivial fraction of the total mass, $(1-\Phi) M$, and the rest are masses which are much smaller than $(1-\Phi) M$. In gelling kernels, collisions between large masses are dominant. After gelation occurs, the smaller masses are consumed by the large mass. For a collision kernel of the form $K(m_1,m_2)\approx \lambda (m_1 m_2)^{\frac{\beta}{2}}$, the criteria for gelation has been established as $\beta>\frac{1}{2}$~\cite{van1985dynamic,connaughton2004stationary}, where $\beta=\nu+\mu$. Figure~\ref{fig:otherkernels} shows the rate function for the collision kernels with $\beta=0,1,2$. The algorithm is able to obtain the rate function for small and large arguments, showing that  a numerical analysis similar to the constant kernel can be done for any arbitrary kernel.
	
In addition to obtaining the large deviation function for the well-known constant, sum and product kernels, we also demonstrate the usefulness of the algorithm by determining the large deviation function for a collision kernel for which the mean field Smoluchowski equation cannot be solved for. Figure~\ref{fig:brownian}  shows the rate function for the Brownian kernel, $K(m_1,m_2)=(m_1/m_2)^{1/3}+(m_2/m_1)^{1/3}+2$, which is widely used in aerosol physics~\cite{veshchunov2010new}.
	
	\subsection{Typical trajectories}\label{typ}
	
	In the algorithm for determining $P(M,N,t)$, the initial condition was fixed as $N(0)=M$, but the final time was varying. Now, we fix the final time to be $T$, \textit{i.e.,} $N(T)=N$, and determine the most probable trajectory under these conditions. We will refer to this trajectory as the instanton trajectory.
	
	To determine the instanton trajectory, we modify the algorithm as follows. The rules to alter the configurations remain the same as before. The rules for assignment of waiting times are modified as follows.  A configuration $1\leq i\leq C$ is chosen. Let the current waiting times associated with the $(i-1)$-th and $i$-th configurations be $\Delta t^\prime_{i-1}$ and $\Delta t^\prime_i$. These waiting times are reassigned, keeping their sum fixed, thus ensuring that the total time taken for $C$ collisions to occur does not change. Let the new waiting times be $\Delta t_{i-1}$ and $\Delta t_i$. Then, $\Delta t_{i-1}$ is drawn from the distribution
	\begin{equation}
	P(\Delta t_{i-1})=\mathcal{N}\mathcal{R}_{i-1}e^{-\mathcal{R}_{i-1}\Delta t_{i-1}}\mathcal{R}_{i}e^{-\mathcal{R}_{i}\Delta t_{i}},
	\end{equation}
	where $\mathcal{N}$ is the normalizing factor, and $\Delta t_i$ is fixed by
	\begin{equation}
	\Delta t_{i}=\Delta t^\prime_{i-1}+\Delta t^\prime_i-\Delta t_{i-1}.
	\end{equation}
	Integrating over $\Delta t_{i-1}$ from $\Delta t_{i-1}=0$ to $\Delta t_{i-1}=\Delta t^{\prime}_{i-1}+\Delta t^{\prime}_{i}$, we obtain 
	\begin{equation}
\mathcal{N}=\frac{(\mathcal{R}_{i-1}-\mathcal{R}_{i})e^{\mathcal{R}_i(\Delta t^{\prime}_{i-1}+\Delta t^{\prime}_i)}}{\mathcal{R}_{i-1}\mathcal{R}_i(1-e^{-(\mathcal{R}_{i-1}-\mathcal{R}_{i})(\Delta t^\prime_{i-1}+\Delta t^\prime_i)})}.
	\end{equation}
	Hence, the final distribution is
	\begin{equation}
	P(\Delta t_{i-1})=\frac{(\mathcal{R}_{i-1}-\mathcal{R}_{i})e^{-(\mathcal{R}_{i-1}-\mathcal{R}_{i})\Delta t_{i-1}}}{1-e^{-(\mathcal{R}_{i-1}-\mathcal{R}_{i})(\Delta t^\prime_{i-1}+\Delta t^\prime_i)}}.
	\end{equation}

To benchmark our simulations, we first ask how the typical trajectories look like. The trajectory obtained for a given $M$ and $\Phi$, without any constraints on the final time and in the absence of bias, is the typical trajectory. We expect that this typical trajectory is described by the Smoluchowski equation (see Sec.~\ref{2}).
	Summing over $i$ and dividing by $M$ in Eq.~(\ref{smol}), gives the rate of decay of the fraction of particles, $n(t)=N(t)/M$ with time.
	
	For the constant kernel $K(m_1,m_2)=\lambda$,
	\begin{equation}
	\frac{dn}{dt}=-\frac{M\lambda n^2}{2}.
	\end{equation} 
	The solution of this equation, with the initial condition $n(0)=1$ is
	\begin{equation}
	n(t)=\frac{1}{1+\frac{M\lambda t}{2}},~ ~\mathrm{constant~kernel}.\label{exact}
	\end{equation}
	This solution describes a typical trajectory provided the number of particles are not of order $1$, which is when the Smoluchowski equation breaks down.
For the sum kernel, $K(m_1,m_2)=\frac{\lambda}{2}(m_1+m_2)$ and the product kernel, $K(m_1,m_2)=\lambda m_1 m_2$, the solution for   the Smoluchowski equation is easily obtained, and are given by~\cite{leyvraz2003scaling}
	\begin{align}
	n(t)=&e^{-\frac{M\lambda t}{2}},~~\mathrm{sum~kernel},\label{sum}\\
	n(t)=&1-\frac{\lambda Mt}{2},~~\mathrm{product~kernel}.\label{prod}
	\end{align}
	We note that these solutions are valid before gelation, where an infinite mass forms in finite time.
Given $\Phi=N/M$, the typical times $t_{typ}$ for the different kernels are obtained by  equating $n(T)$ in Eqs.~(\ref{exact}), (\ref{sum}) and (\ref{prod}) to $\Phi$. To check that the simulations reproduce the typical trajectories, we set $T=t_{typ}$, and then ask whether the numerically obtained instanton solution matches with the solution to the Smoluchowski equation.

\begin{figure}
		\centering
		\includegraphics[width=\columnwidth]{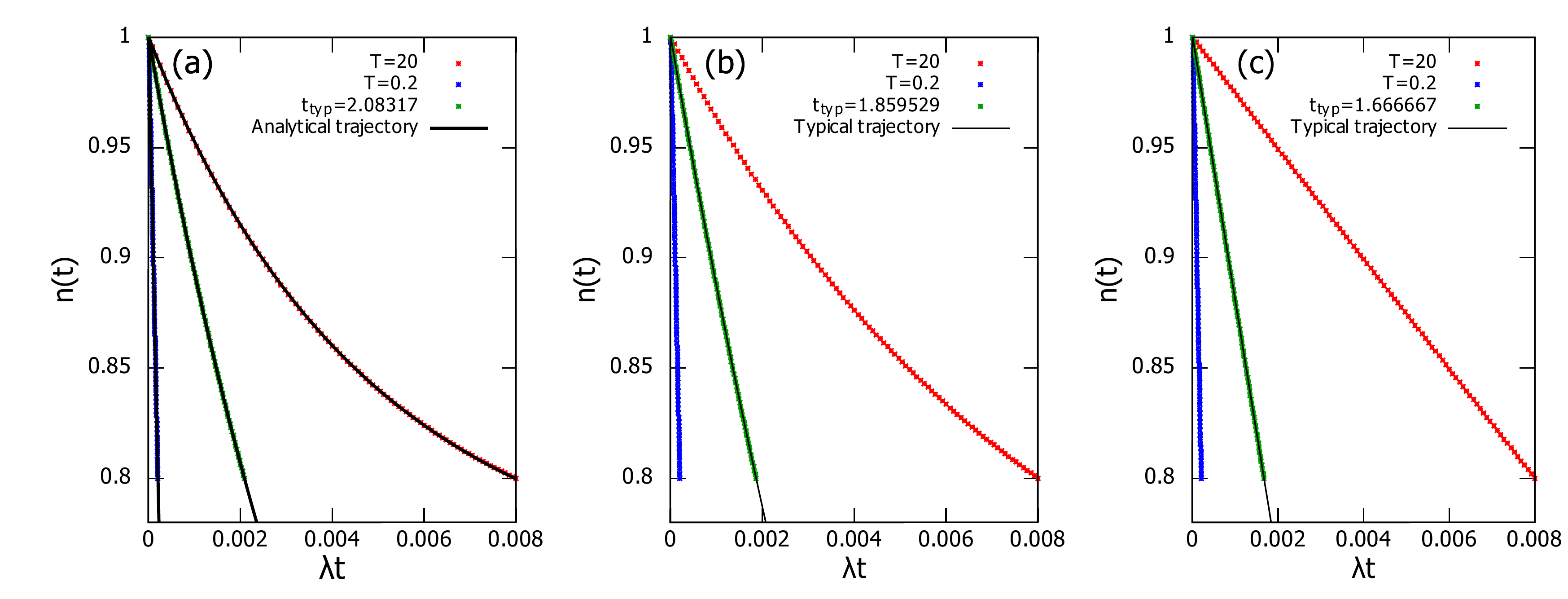}
		\caption{Instanton trajectories for different final times $T$
		are shown for (a) constant kernel and $\Phi=0.3$, (b) sum kernel and $\Phi=0.5$, and (c) product kernel and $\Phi=0.8$. The three times shown are for $T=t_{typ}$, $t_{typ}/10$ and $4 t_{typ}$.  The data for the typical times are compared with the exact solution of the Smoluchowski equation [see Eqs.~(\ref{exact}), (\ref{sum}) and (\ref{prod})]. For the constant kernel, the data for atypical times are also compared with the exact result [see Eqs.~(\ref{shorttime}), (\ref{longtime}). The data are for $M=240$.}
		\label{fig:instanton}
	\end{figure}

Figure~\ref{fig:instanton} shows the numerically obtained instanton trajectories for the constant, sum, and product kernels, for typical as well as atypical final times $T$, for $\Phi=0.8$. For $T=t_{typ}$, the data are in excellent  agreement with the solution of the Smoluchowski equation for all the three kernels, thus providing a check for the correctness of the implementation of the algorithm. The algorithm is also able to obtain the instanton trajectories for atypical trajectories for times which are both much smaller than as well as much larger than the typical times. The exact answers for the atypical trajectories of the constant kernel are~\cite{subashrildcca}
\begin{align}
n(t)&=-p\tan\frac{M\lambda p(t-t_0)}{2}~~~~~T<t_{typ},\label{shorttime}\\
n(t)&=q\coth\frac{M\lambda q(t-t_1)}{2}~~~~~T>t_{typ}\label{longtime},
\end{align}
where $p$, $t_0$, $q$ and $t_1$ are determined from the boundary conditions $n(0)=1$ and $n(T)=\Phi$. The simulation results are in excellent agreement with the exact results for the instanton trajectories for atypical events, as shown in Fig.~\ref{fig:instanton}.

We also check that the minimum of the large deviation rate function for the constant, sum, and product kernels for $\Phi=0.8$, shown in Fig.~\ref{fig:otherkernels}, occurs at the typical times as calculated by the Smoluchowski equation in Eqs.~(\ref{exact}), (\ref{sum}), and (\ref{prod}).

	\subsection{Autocorrelation times}\label{autocorr}
	To characterize the algorithm, we determine the dependence of the autocorrelation time on bias $w$, fraction $\Phi=N/M$ and the parameter $p$. We recall that $p$ is probability that in a given micro-step the sequence of collisions is modified, while $(1-p)$ is the probability that the waiting times are modified. The autocorrelation function, $ACF(\tau)$, for a stationary variable $X$ is defined as 
	\begin{equation}
	ACF(\tau)=\frac{1}{T^\prime}\int_{0}^{T^\prime}dt [X(t+\tau)-\braket{X}][(X(t)-\braket{X})],
	\end{equation}
	where $T^\prime$ is the total time over which  $X$ is measured, and $\tau$ is the delay. 
	
	The Monte Carlo algorithm involves introducing local modifications to the trajectory by changing either the waiting time associated with a collision or the sequence of collisions. To measure the  autocorrelation in time as well as in configuration space, we define
	\begin{align}
	t=&\sum_{i=0}^C \Delta t_i, \\
	Q_i=&\sum_{m=1}^M m^2 N_i(m),~~i=2,\ldots, C.\label{Q}
	\end{align}  
$Q_i$ is a measure of the mass distribution after the $i$-th collision. 	We choose the second moment of mass, as it is the lowest moment that changes when the mass distribution is modified, the zeroth and first moments being constants. 

\begin{figure}
	\centering
	\includegraphics[width=0.85\columnwidth]{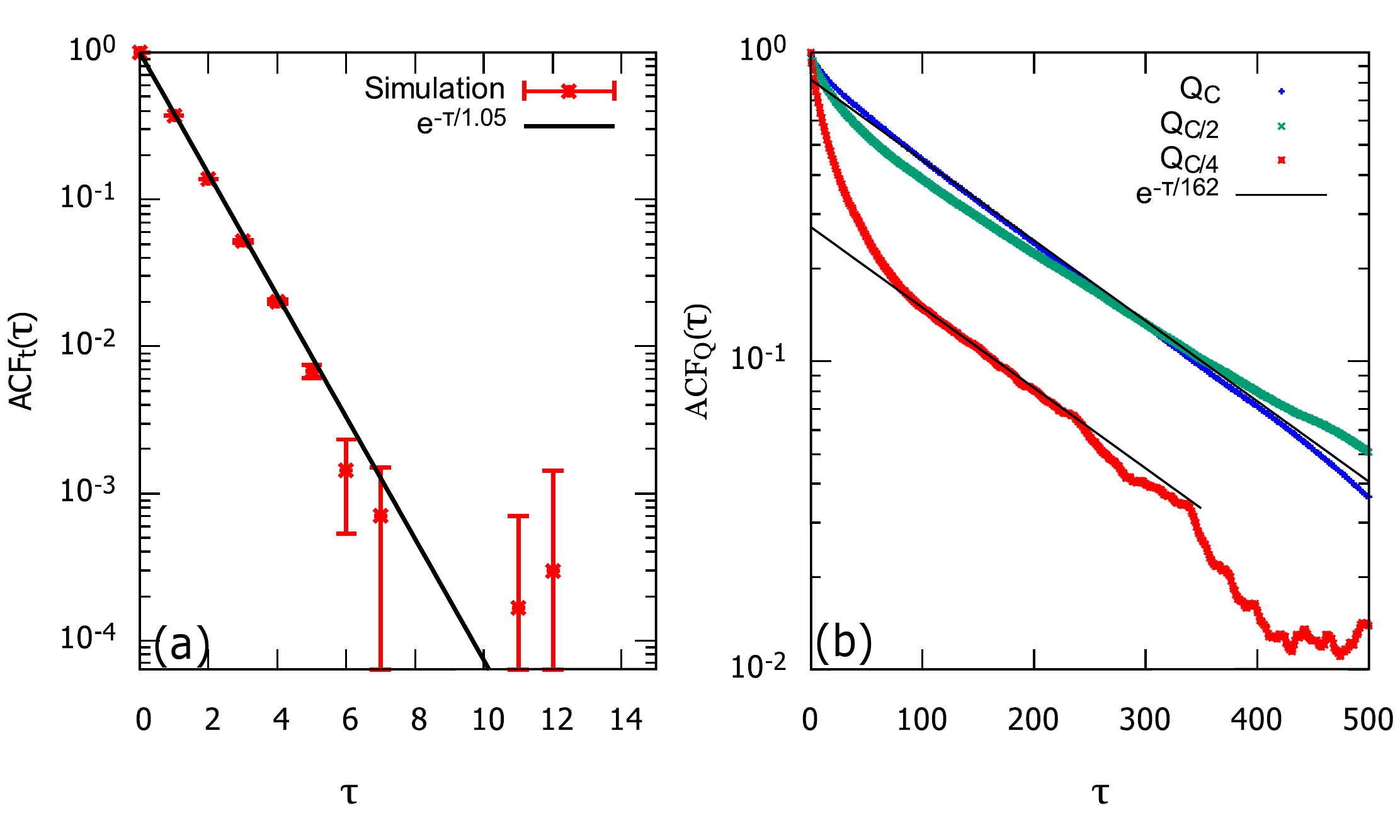}
	\caption{Autocorrelation functions for (a) total time $t$ and $C=60 $, and (b) $Q_i$, for $i=C,C/2,C/4$, where $C=24$, with delay $\tau$, for the constant kernel, for $M=120, w=0, p=0.5$.}
	\label{fig:autocorr-fn}
	\end{figure}

The autocorrelation functions $ACF_t (\tau)$ and  $ACF_Q (\tau)$, corresponding to $t$ and $Q$ decay exponentially with time, as shown in Fig.~\ref{fig:autocorr-fn}. To decide which configuration we should use for the $Q$ autocorrelation, we compare the  autocorrelation functions for the $C$-th, $C/2$-th, and $C/4$-th collisions in Fig.~\ref{fig:autocorr-fn} (b). We find that the correlation time, determined by the slope of the curve on the semi-log plot, is nearly the same for all the three data. For convenience, we choose the $C$-th configuration, henceforth, to measure the autocorrelation time $\tau_Q$, and will drop the subscript $i$ from the second moment $Q$ in Eq.~(\ref{Q}).
We define autocorrelation times, $\tau_t$ and $\tau_Q$ via 
	\begin{align}
	\frac{ACF_t(\tau)}{ACF_t(0)}\approx& e^{-\tau/\tau_t},\\
	\frac{ACF_Q(\tau)}{ACF_Q(0)}\approx& e^{-\tau/\tau_Q}.
	\end{align}
The autocorrelation times $\tau_t$ and $\tau_Q$ are obtained by fitting these exponential functions to the exponentially decaying regions of $ACF_t(\tau)$ and $ACF_Q(\tau), $ respectively.
We now characterize the dependence of $\tau_t$, $\tau_Q$ on the fraction of particles remaining, $\Phi$, bias $w$, and the parameter $p$. All the simulations have been performed for the constant kernel.
	
\begin{figure}
		\centering
		\includegraphics[width=0.85\columnwidth]{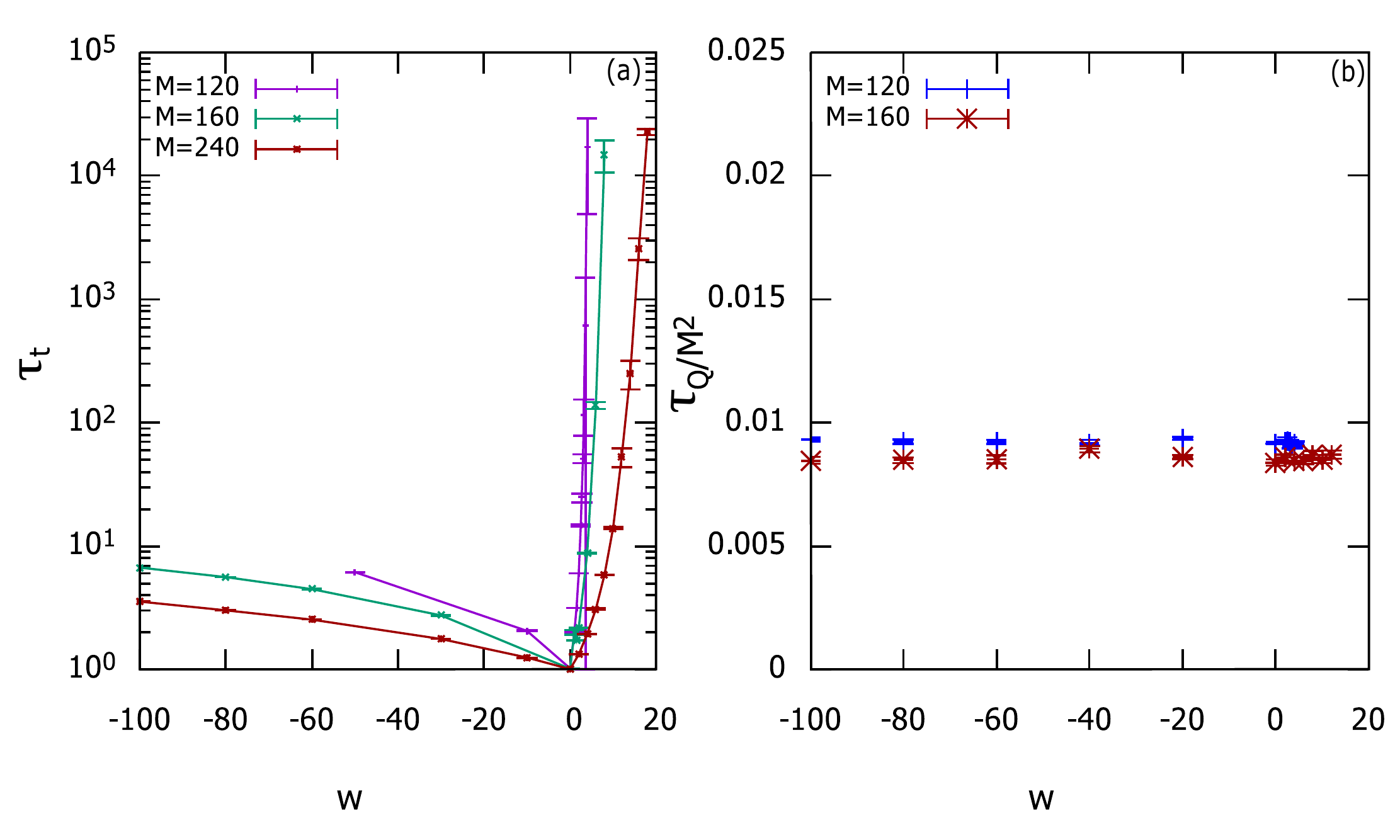}
		\caption{The variation of the autocorrelation times (a) $\tau_t$ and (b)$\tau_Q/M^2$ with $w$ for different $M$. The data are for the constant kernel for $p=0.5$ and $\Phi=0.8$.}
		\label{fig:autocorr-tau-w}
	\end{figure}

	Figure 8 shows the dependence of $\tau_t$ and $\tau_Q$ on the bias, $w$ for fixed $\Phi=0.8$ and $p=0.5$. For $w>0$, $\tau_t$ increases sharply with $w$ and diverges at the cutoff bias [see Fig.~\ref{fig:autocorr-tau-w} (a)]. For $w<0$, $\tau_t$ increases much more slowly. We find that $\tau_t$ decreases with $M$, however, we cannot find a scaling behaviour. For the unbiased case, $w=0$, we find that $\tau_t$ is independent of $M$. In contrast, we find that $\tau_Q$ shows at most a very weak dependence on $w$. It increases with $M$, but the data for different $M$ collapse onto one curve when $\tau_Q$ is scaled by $M^2$ [see Fig.~\ref{fig:autocorr-tau-w}(b)].

\begin{figure}
		\centering
		\includegraphics[width=0.85\columnwidth]{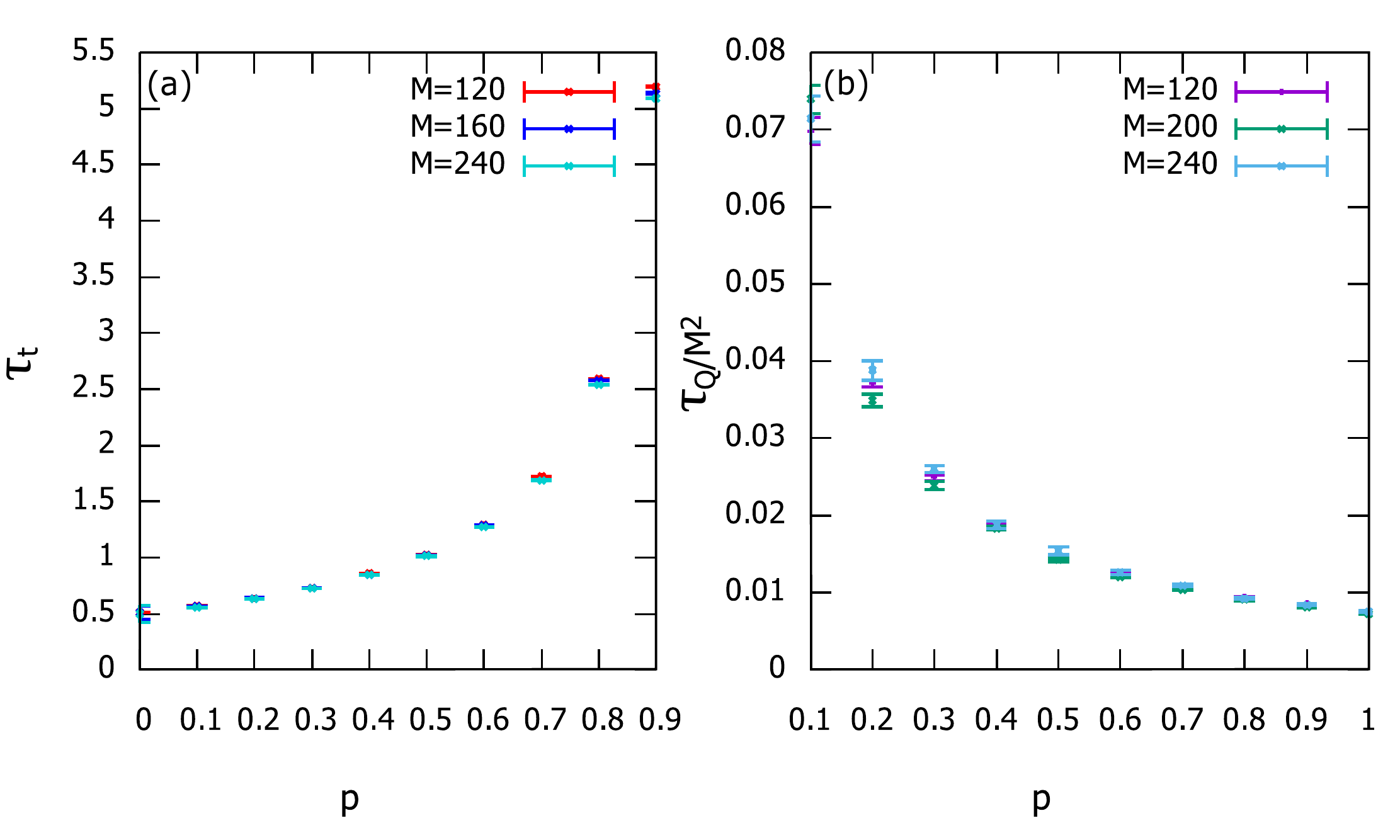}
		\caption{The variation of the autocorrelation times (a) $\tau_t$ and (b)$\tau_Q/M^2$ with $p$ for different $M$. The data are for the constant kernel for $\Phi=0.8$ and $w=0$.}
		\label{fig:autocorr-tau-p}
	\end{figure}

The variation of $\tau_t$ and $\tau_Q$  with the parameter $p$ is shown in Fig.~\ref{fig:autocorr-tau-p} for fixed $\Phi=0.8$ and $w=0$. $\tau_t$ diverges as $p\to 1$. This is expected since, in this limit, the probability of modifying waiting times tends to zero. We also find that  $\tau_t$ is independent of $M$. $\tau_Q$, on the other hand, increases with $M$. However, the data for different $M$ collapse onto one curve when $\tau_Q$ is scaled by $M^2$. As expected, $\tau_Q$ diverges for small $p$ because the probability of updating configurations tends to zero in this limit.

We also checked the variation of $\tau_t$ and $\tau_Q$ with the parameter $p$ for non-zero values of $w$. We again find that the data for $\tau_Q$ collapse when scaled by $M^2$. However, we do not find a scaling for $\tau_t$.

\begin{figure}
	\centering
	\includegraphics[width=0.85\columnwidth]{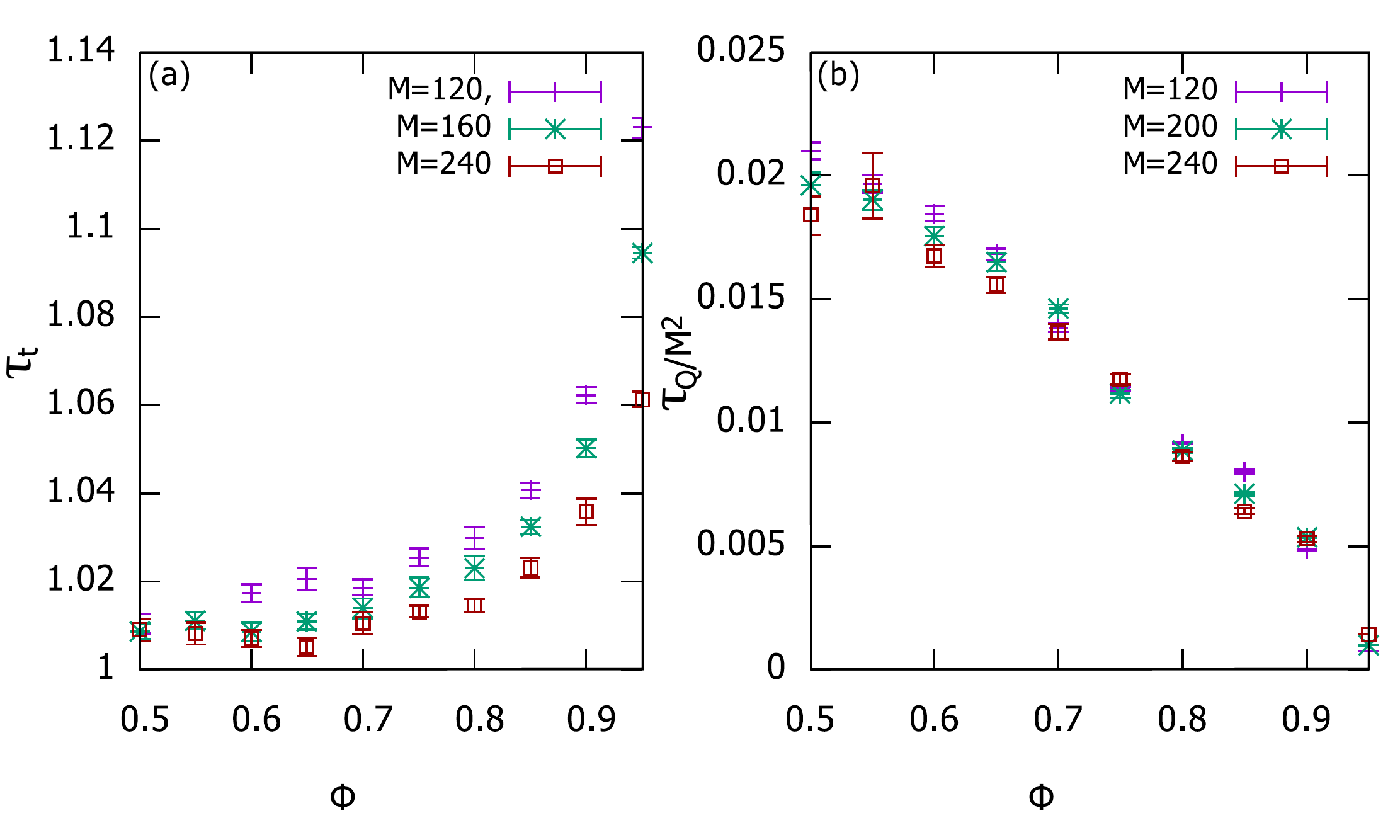}
	\caption{The variation of the autocorrelation times (a) $\tau_t$ and (b)$\tau_Q/M^2$ with $\Phi$ for different $M$. The data are for the constant kernel for $p=0.5$ and $w=0$.}
	\label{fig:autocorr-tau-phi1}
\end{figure}

The variation of $\tau_t$ and $\tau_Q$  with $\Phi$ is shown in Fig.~\ref{fig:autocorr-tau-phi1} for fixed $p=0.5$ and $w=0$. $\tau_t$ is order $1$ and very weakly dependent on both $\Phi$ as well as $M$. For $\tau_Q$, like before, the data for different $M$ collapse onto one curve when $\tau_Q$  is scaled by $M^2$. We also find that $\tau_Q$ is larger for smaller $\Phi$.
	
From Figs.~\ref{fig:autocorr-tau-w}-\ref{fig:autocorr-tau-phi1}, we see that $\tau_t$ remains small unless $p\to1$, or if the positive bias is close to the cutoff bias. On the other hand, $\tau_Q$ is order of $M^2/100$ times larger than $\tau_t$. Choosing a value of $p$ close to $1$ will optimize the implementation of the algorithm, keeping both autocorrelation times finite. 

	\section{Summary and Conclusion \label{sec:summary}}
	
	To summarize, we developed a biased Monte Carlo algorithm to compute probabilities of rare events in irreversible cluster-cluster aggregation for an arbitrary collision kernel. In particular, the algorithm measures $P(M, N, t)$, the probability of $N$ particles remaining at time $t$ when there are $M$ particles initially, as well as the most probable trajectories for fixed $M$, $N$, and $t$. By choosing appropriate biases, the algorithm can efficiently sample the tails of the distribution with low computational effort. We prove that the algorithm is ergodic by specifying a protocol that transforms any given trajectory to a standard trajectory using valid Monte Carlo moves. The algorithm is benchmarked against the exact solution for the constant kernel. 
	
	To characterize the algorithm, we define autocorrelation times $\tau_t$ and $\tau_Q$,  corresponding to the waiting times as well as the configurations. We find that $\tau_t$ is much smaller than $\tau_Q$ for almost the entire range of parameters. From simulations for different $M$, we find that  $\tau_t$ is at most only weakly dependent on $M$, while  $\tau_Q$ is proportional to $M^2$. Based on the dependence of $\tau_t$ and $\tau_Q$ on the bias $w$, the fraction of particles remaining $\Phi=N/M$, and the parameter $p$ which decides what fraction of the Monte Carlo moves are changes to configurations, we conclude that it is best to choose a value of $p$ as close to $1$ as possible.
	
	Generalizing the numerical results for constant, sum, and product kernels, we conclude that there exists a large deviation principle for arbitrary kernels, where the total mass $M$ is the rate.  This provides hints for a more rigorous treatment of the large deviation function for the problem of aggregation. In a future publication, based on the insights gained from this paper, we will provide a derivation of the large deviation function for some kernels. 		
	
	Although this paper deals with binary aggregation, the algorithm that we have developed can also be easily generalized to the numerical study of the non-binary processes $k A\rightarrow \ell A$, with suitably modified rates. Adding spatial degrees of freedom, and transport, like diffusion, is a problem of interest. However, generalizing the algorithm to such systems is a challenging problem. Adding a competing process such as fragmentation is another problem of interest~\cite{brilliantov2015size,connaughton2018stationary,brilliantov2021nonextensive,kalinov2022direct}.Competing processes like these can lead to phase transitions and oscillations, at least in the mean field limit~\cite{ball2012collective}. These are promising areas for future study. 

{\em Author contributions:} V. Subashri developed and implemented the algorithm, and wrote the paper. R. Dandekar was involved in the development of algorithm. R. Rajesh and O. Zaboronski conceived and directed this work, and helped in writing the paper.

	\bibliographystyle{elsarticle-num}

\end{document}